# Direct observation of a localization transition in quasi-periodic photonic lattices

Y. Lahini[1], R. Pugatch[1], F. Pozzi[2], M. Sorel[2], R. Morandotti[3], N. Davidson[1] and Y. Silberberg[1]

[1]*Department of Physics of Complex Systems, the Weizmann Institute of Science, Rehovot, Israel*

[2]*Department of Electrical and Electronic Engineering, University of Glasgow, Glasgow, Scotland*

[3]*Institute National de la Recherche Scientifique, Varennes, Québec, Canada*

**The localization of waves in non-periodic media is a universal phenomenon, occurring in a variety of different quantum and classical systems, including condensed-matter [1-3], Bose-Einstein condensates in optical lattices [4, 5], quantum chaotic systems [6, 7], sound waves [8] and light [9-18]. A localization phase transition is expected to occur in three dimensional disordered systems as the strength of disorder crosses a critical value [1, 3]. Recently, a crossover from an extended to a localized phase has been observed in low-dimensional photonic lattices [16-18] and Bose-Einstein condensates [4, 5]. Other experiments studied the critical behaviour near the transition in three dimensions via transmission measurements [14, 15]. However, no direct observation of a localization transition for light has been reported. In 1979 Aubry and André predicted that for a certain class of quasi-periodic potentials, a localization phase transition can occur already in one-dimension [19]. This strongly correlated potential is markedly different from the disordered case were the potential is uncorrelated. Here we report an experiment that realizes the Aubry-André model in quasi-periodic photonic lattices. We observe the signature of a localization phase transition by directly measuring the expansion rates of initially narrow wave**

**packets propagating in the lattice. Below the transition point, all the modes of the system are extended and therefore an initially narrow wave-packet eventually spreads across the entire lattice. Above the critical point, all modes are localized and expansion is suppressed. In addition, we study the effect of weak nonlinear interactions on light propagation below and above the transition.**

The Aubry-André model (AA) describes a quasi-periodic lattice composed of a 1D periodic lattice that is modulated by a frequency that is maximally incommensurate with respect to the lattice period [19]. This model was originally described in the tight-binding formulation:

$$i\frac{\partial \psi_n}{\partial t} = H_{AA}\psi_n = [\beta_0 + \lambda C \cos(2\pi n\chi)]\psi_n + C(\psi_{n-1} + \psi_{n+1}) \quad (1)$$

where $\psi_n$ is the wavefunction at site n, $\beta_0$ is the single-site energy of the (unperturbed) periodic lattice, C is the tunneling rate between sites, $\chi = (\sqrt{5}+1)/2$ and is the golden mean. The Aubry-Andre Hamiltonian ($H_{AA}$) has an n↔k duality point at a modulation strength λ=2, where n is the lattice site index and k is the fractional Fourier number: k=2πχ/n [19,20] Consequently, a localization phase transition takes place at this modulation strength (At this point, the Aubry-André Hamiltonian coincides with the Harper Hamiltonian [21] that exhibits the famous Hofstadter butterfly fractal spectrum). In Fig. 1a we plot the amplitude of the ground state as a function of λ, calculated using Eq. 1 for a system with 99 sites. At λ=0 $H_{AA}$ is periodic and as a result the ground state is extended over the entire lattice. As λ is increased the ground state becomes rugged, but remains extended. Right above λ=2 the ground state exhibits a sharp localization transition, and becomes localized around a single lattice site. All other eigenmodes of $H_{AA}$ exhibit the same sharp transition, becoming tightly localized at different sites in the lattice above λ=2. For comparison we present the same calculation for a disordered system based on the Anderson model [1] (Fig 1b):

$$i\frac{\partial \psi_n}{\partial t} = H_A \psi_n = [\beta_0 + \lambda C W_n]\psi_n + C(\psi_{n-1} + \psi_{n+1}) \qquad (2)$$

where $W_n$ are random numbers drawn from a flat distribution ranging from -1 to 1. In this case there is no sharp localization transition at finite $\lambda$. Instead, there is a smooth decrease of the localization length as the strength of the disorder $\lambda$ is increased. For infinite systems, all eigenmodes of the Anderson Hamiltonian ($H_A$) are localized, even for arbitrarily small disorder strengths [3]. In a finite size system and weak enough disorder, the eigenstates appears to be extended since their localization length becomes larger than the system size. As disorder increases, the eigenmodes become localized within the finite lattice. This, however, cannot be considered as a phase transition since the amount of disorder needed to observe localization goes to zero as the system's size goes to infinity. In contrast, in the Aubry-André model all eigenmodes becomes localized above $\lambda_c=2$, with a localization length which is smaller than the size of the system. The width of the critical region around $\lambda_c=2$ only weakly depends on the system size, and the transition is clearly visible for a system with 99 sites (see Fig 1a).

The localization transition has a profound effect on the transport properties of the system [22]. For example, consider a spatially narrow initial condition confined to a single site. Such a wave-packet will expand indefinitely if the eigenmodes of the system are extended, but will come to a halt if all the eigenmodes are localized. In this work we employ this physical indicator for localization and observe a localization transition for expanding wavepackets in one-dimensional photonic lattices. A recent experiment by Roati et. al. on localization of ultra-cold atoms with reduced interaction strength [5], implemented a related Hamiltonian with a continuous potential which is a sum of two sine waves with a frequency ratio of $\chi=1.1972$ [5]. The authors reported a smoothed transition at values of $\lambda >7$, where they observes localized states every fifth site.

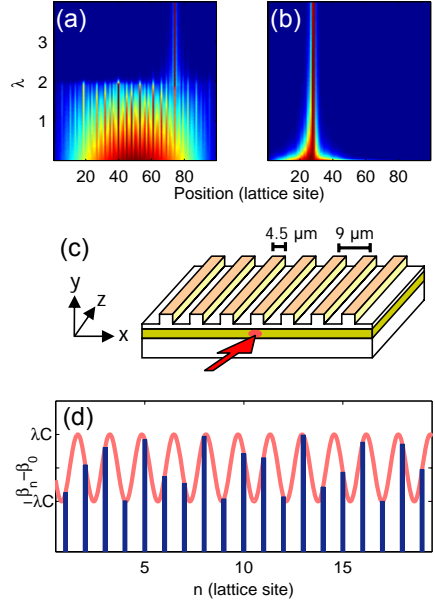

Fig. 1. (a) Calculated amplitude profile of the ground state of an incommensurate lattice (horizontal axis) versus the strength of the incommensurate modulation depth λ (vertical axis). A sharp localization transition is observed at λ=2. The wavefunction is extended for λ<2, and localized to a single site for λ>2. A similar transition occurs for all other eigenstates of the system. (b) The same calculation for a disordered lattice. Here the localization length goes continuously to zero. All horizontal cross sections are normalized to a unit maximum. (c) A schematic view of the photonic lattice used in the experiments. (d) A quasi-periodic lattice is obtained by modulating the on-site eigenvalues ($β_n$ in Eq. 1) of the lattice sites at a frequency that is incommensurate with the frequency of the lattice.

In our experiment we use a one-dimensional photonic lattice with a carefully designed refraction index step pattern that forms a lattice of evanescently coupled waveguides (see Fig 1.c) [23]. Structures of this type have been recently used to study linear and nonlinear phenomena in one and two dimensional disordered lattices [16-18], featuring the observation of localized eigenmodes and exponential (Anderson) localization of expanding wave packets, and enabling the study of the interplay between nonlinearity and disorder. Similarly, two-dimensional Photonic quasi-crystals were realized and studied in [24]. The important feature of these lattices is that the time evolution of waves inside the lattice is now combined with the propagation along the z direction. The experimental advantage of this approach is that it is possible to construct exact initial conditions, by selecting the input position of the initial wave packet, and then observe its evolution by measuring the light distribution at the output. This way, a direct observation of the transport properties of the lattice can be made, in contrast to traditional transmission or conductance measurements. Another important advantage is the ability to introduce nonlinearity in a controlled manner and directly observe its effect on transport.

The propagation of light in such lattices has been shown to be well described by the discrete nonlinear Schrödinger equation [23]. For an incommensurate photonic lattice, Eq. 1 will now read:

$$i\frac{\partial \psi_n}{\partial z} = [\beta_0 + \lambda C \cos(2\pi n \chi)]\psi_n + C(\psi_{n-1} + \psi_{n+1}) + \gamma |\psi_n|^2 \psi_n \quad (3)$$

Where z is the free propagation axis z=ct, c being the speed of light in the medium, $\beta_0$ is the single site propagation eigenvalue of the underlying periodic lattice, and $\gamma$ is the Kerr nonlinear coefficient. In the linear limit ($\gamma=0$) the equations are identical to the tight-binding model of solid state physics, while for finite nonlinearity these equations are also used to describe the mean-field evolution of an interacting Bose-Einstein condensates in an optical lattice [25].

To realize the Aubry-Andre model, we fabricated a set of quasi-periodic waveguide lattices, with an incommensurate ratio $\chi = (\sqrt{5}+1)/2$ of and different values of $\lambda$ (see methods). To measure the expansion of the wave packets, we injected a narrow beam into a single site at the input facet of the lattices, and measured the output intensity distribution. For each lattice we have repeated this procedure for up to 60 different sites in the same lattice and averaged the result. Identical samples with different lengths were used to directly measure the expansion rate of the wave packets. Results are presented in Fig 2. Figure 2a depicts the expansion of a single-site wave packet in a periodic lattice (λ=0) for two different times (equivalent to 6mm and 21mm of propagation). Fig 2b depicts the same results for an incommensurate lattice with λ=1.6. Here too the wavepackets expand throughout the propagation, yet at a reduced speed, as expected from theory [22]. When an incommensurate lattice with λ=3.1 was used (Fig. 2c), the wave packet remained tightly localized to the input site, and did not expand throughout the propagation. The width of a wave packet can be characterized by the participation ratio (PR) for the wavefunction, given by:

$$PR = \frac{\left(\sum |\psi_n|^2\right)^2}{\sum |\psi_n|^4}$$

A measure of the transition can be obtained by plotting the participation ratio along the propagation for different values of λ. The results are presented in Fig 2d. This figure shows the transition from expansion to localization as a function of the incommensurate modulation strength. The collapse of the curves for λ>2 as well as their saturation indicate the localization transition. In order to compare our results to theoretical predictions, we plot the measured PR of the wavepackets after 21mm of propagation and the expected values calculated using Eq. 3.

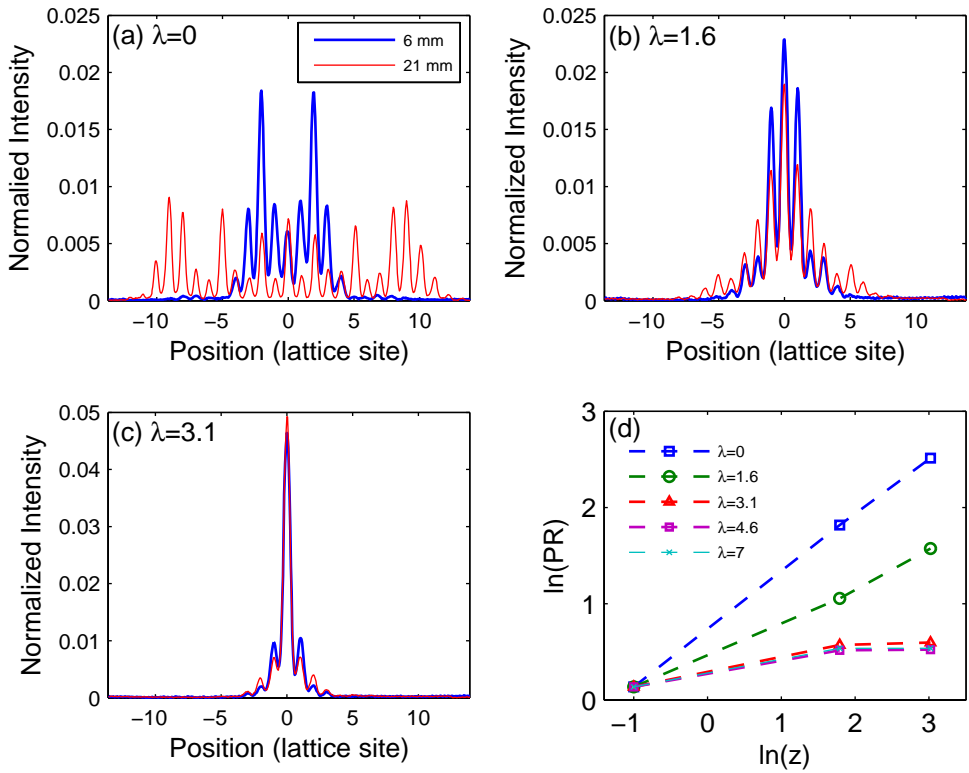

Fig. 2 Experimental measurements of a localization transition in incommensurate photonic lattices with 99 lattice sites. (a) The expansion of a single site initial wave packet in a perfectly ordered lattice (λ=0) after 6 mm of propagation (blue) and 21 mm (red). (b) The same for an incommensurate lattice with λ=1.6. The wavepacket still exhibits expansion, but at a reduced speed. (c) The same for an incommensurate lattice with λ=3.1. The wavepacket remains tightly localized about the input position throughout the propagation, signifying localization. (d) Logarithmic plots of the averaged width of the wavepackets (measured by the participation number PR) at different lengths, for different λ values. This figure shows the transition from expansion to localization as a function of the incommensurate modulation strength, signifying the localization transition.

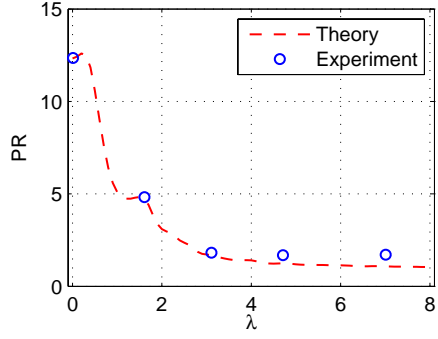

Fig. 3 Comparison between the measured width of the wavepackets after 21mm of propagation (dots) and the theoretical values expected from Eq. 3 (dashed line), for different values of λ.

    Since the photonic lattices used in these experiments display nonlinear effects at high light intensities, they offer the possibility to study experimentally the effect of nonlinear interaction on the propagation before and after the transition. The role of nonlinear interactions in incommensurate lattices was considered theoretically in [26-28]. An experimental investigation was made of BEC in incommensurate lattices [29], yet the inherently strong interactions in this case are expected to result in a different quantum phase [30] which cannot be described by Eq. 3. The measured effect of nonlinearity on wave packet expansion in our system is presented in Fig 4a. Here we plot the averaged width of the wavepackets after 21mm of propagation as a function of nonlinearity, for different incommensurate modulation strengths λ. As the results show, for λ=0 an increased nonlinearity results in a reduced expansion of the wave packet. This is expected [23], since the nonlinearity in our system is positive ($\gamma>0$), corresponding to attractive or self-focusing interaction. For λ>2 (the localized regime) we find that at some of the input positions, wavepackets that are localized in the linear regime exhibit weak expansion under the influence of nonlinearity. A typical example is presented in Fig. 4b for a lattice with λ=3.1. This effect underlies the increase in the averaged participation ratios presented in Fig 4a for λ>2, indicating a slight increase of the localization length induced by nonlinearity. For λ=1.6 (still below the

localization transition) the expansion is reduced at low powers, and slowly recovers at higher powers. We note that the increase of the participation number with nonlinearity is not reproduced by simulations based on Eq. 3. Possible mechanisms for this discrepancy are higher order nonlinear effects or nonlinear absorption terms.

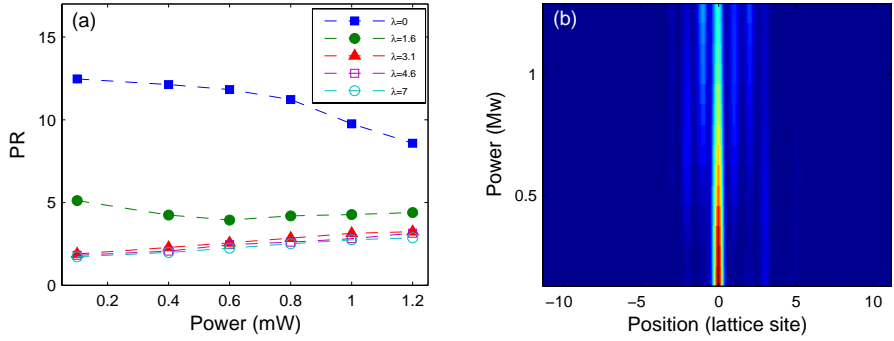

Fig. 4 Experimental measurements of the effect of nonlinearity on wave packet expansion in incommensurate photonic lattices. (a) Plots of the averaged participation ratios measured after 21mm of propagation, as a function of power and for different incommensurate modulation depths λ. For λ=0, an increased nonlinearity results in a reduced expansion of the wave-packet. However for λ>2 (the localized regime) nonlinearity results in weak expansion. (b) The measured output intensity distribution (horizontal axis) versus the input beam power for a single (non-averaged) initial condition, in a lattice with λ=3.1. In the linear case the output is tightly localized to the input site, however, it spreads significantly as the power is increased.

In conclusion, we have realized the Aubry-André model for localization using a one dimensional lattice with an incommensurate modulation. Evidence for a localization phase transition expected to occur in such lattices was obtained by measuring the spreading of initially narrow wavepackets in lattices with different incommensurate modulation strength. Below the transition point we found that wavepackets that couples to a single input site expands in the lattice, while above the transition point the expansion of the wave packets was brought to a halt. By introducing nonlinearity in a controlled manner, we have demonstrated that in the localized regime, nonlinear effects result in a slight increase in the width of the localized wavepackets. Possible extensions of this work involves the scrutiny of the critical regime around λ=2 and the study of the role of nonlinearity near the localization transition.

**Methods**

To realize the Aubry-André model, a set of quasi-periodic waveguide lattices on an AlGaAs substrate was fabricated. The samples were prepared using high resolution, large field e-beam lithography (Vistec VB6), followed by reactive ion etching. The width of the waveguides is designed to yield an incommensurate modulation according to Eq. (2) with $\chi = (\sqrt{5}+1)/2$ , the golden mean (see Fig 1.d), where the average waveguide width is 4.5 microns. The tunneling parameter between sites C is determined by the etch depth of the sample and by the distance between neighboring waveguides. This distance was designed such that the tunneling rate is identical for each waveguide pair, with an average waveguide separation of 5 microns. The final values for the coupling rate C was measured directly from the propagation in the periodic lattice. Knowing the design profile of the sample and the coupling, we calculated the etch depth to be 1.3 microns. This value was verified independently using a surface profiler (Zygo NewView 5000). Seven samples were fabricated, with different values of the parameter λ. The obtained modulations rates λ were calculated numerically from the design profile, and from the etch depth of the fabricated sample. The finite resolution of the fabrications process (50 nm) produces a small amount of on-site disorder added to the AA model. Using numerically exact calculations based on the final fabricated profile we were able to estimate the strength of this fabrication-related disorder to be less than 0.05*C. We then verified that this amount of disorder has no significant effect on the localization transition, as the localization length associated with this disorder is much larger than our system.

We would like to thank N. Bar-Gill for illuminating discussions. Financial support by NSERC and CIPI (Canada), and EPRSC (UK) is gratefully acknowledged. YL is supported by the Adams Fellowship of the Israel Academy of Sciences and Humanities.

**Correspondence and requests for materials should be addressed to Y. S.** (e-mail: **yaron.silberberg@weizmann.ac.il**).